\documentstyle[multicol,aps,prl,graphicx,amssymb]{revtex}
\begin{document}

\draft

\title{Zeroth principle of thermodynamics in aging quasistationary states}

\author{Luis G. Moyano, Fulvio Baldovin and 
Constantino Tsallis\thanks{E-mail addresses: moyano@cbpf.br, 
baldovin@cbpf.br, tsallis@cbpf.br}}
\address{Centro Brasileiro de Pesquisas F\'{\i}sicas\\ 
Rua Xavier Sigaud 150, Urca 22290-180
Rio de Janeiro, Brasil}

\date{\today}

\maketitle

\begin{abstract}
We show that the zeroth principle of thermodynamics applies to aging 
quasistationary states of long-range interacting
$N$-body Hamiltonian systems.
We also discuss the measurability of the temperature in
these out-of-equilibrium states
using a {\it short-range} interacting thermometer.
As many connections are already established between such quasistationary
states and nonextensive statistical mechanics, our results
are the first evidence that such basic concepts apply to
systems that the nonextensive formalism aims to describe.
\end{abstract}


\begin{multicols}{2}
The zeroth principle --- if systems $A$ and $B$ are in thermal equilibrium with $C$, they are in thermal equilibrium among them ---, is one of the basic principles in physics. It establishes the transitivity of the temperature, and  
its importance can hardly be overestimated, being essential to the 
logical formulation of thermodynamics. In particular, it is fundamental 
in thermometry, which resides at the very grounds of 
experimental physics.
Recently, considerable attention has been driven to  
$N$-body Hamiltonian systems that constitute paradigmatic models for 
long-range interactions 
\cite{antoni_01,anteneodo_01,latora_01,dauxois_01}.
These systems present several anomalies in their behavior. 
Among them, negative microcanonical specific heat, 
inequivalence between canonical and
microcanonical ensembles, 
vanishing Lyapunov spectrum, L\'evy walks and anomalous diffusion.
Quite remarkably, 
aging has also been exhibited \cite{montemurro_01}.
In this Letter we focus on the appearance of long-lasting 
(possibly infinite-lasting in the thermodynamical limit) 
metastable or quasistationary 
states (QSS), characterized by a non-Gaussian velocity distribution and 
by a temperature that does not coincide 
with the one  predicted by the Boltzmann-Gibbs (BG) theory \cite{latora_02,nobre_01}.
The BG thermal equilibrium is attained at much later times, after a crossover time which diverges with $N$.
Using a standard dynamical definition of the 
temperature, we show that  {\it the zeroth principle of thermodynamics  applies to these QSS in a manner which is essentially the same as for conventional thermal equilibria}. 
Moreover, we exhibit that thermalization occurs even if measured with a 
thermometer that, in contrast with the `thermal bath' that is 
probed, is {\it short-range} interacting.  
As these systems present a considerable number of connections 
\cite{anteneodo_01,latora_01,montemurro_01,latora_02,nobre_01,tsallis_01} 
with nonextensive statistical mechanics  \cite{tsallis_02}, 
our findings constitute a strong suggestion
that these basic concepts apply to situations 
that the nonextensive formalism aims to describe.

Long-range interacting systems constitute nowadays an exciting 
frontier topic in many areas of physics (astrophysics, nuclear
physics, plasma physics, Bose-Einstein condensates, atomic clusters,
hydrodynamics, among others) \cite{dauxois_02}.
They also provide
an interesting arena in a trans-disciplinary perspective as prototypical
systems that enable the study of 
analogies or differences between alternative approaches. 
In particular, interesting results are now available which exhibit
\cite{dauxois_01,latora_02,nobre_01}
inequivalences between standard BG
approaches and methods of dynamical systems. 
If we consider, as Einstein pointed out in 1910 \cite{cohen_01},
that the foundations of statistical mechanics lies on dynamics,
this is a major point worthwhile to be deeply analyzed.

As a representative example of such a richness of behavior, 
the Hamiltonian Mean Field model, which describes a system of 
$N$ planar classical spins interacting through an infinite-range 
potential, has been largely considered in the literature 
\cite{antoni_01,anteneodo_01,latora_01,dauxois_01,montemurro_01,latora_02}.
This Hamiltonian can be written as
\begin{equation}
H = K+V=\sum_{i=1} ^{N} \frac{p_{i}^2}{2} + 
\frac{1}{2N} \sum_{i,j=1} ^{N} 
\left[1-cos(\theta_{i}-\theta_{j})\right],
\label{H}
\end{equation}
where $\theta_i$ is the $i$th angle and $p_i$ is the conjugate variable 
representing the angular momentum (unit inertial moment is assumed).  It is the 
inertial version of the $XY$ ferromagnetic spin model, with the interaction terms connecting not only first neighbors but all couples. Note that it is common use,
although not necessary \cite{anteneodo_01}, to divide the potential 
term by $N$ in order to make the Hamiltonian formally extensive. 
Defining the mean field vector 
${\mathbf M}\equiv \sum_{i=1}^N{\mathbf m}_i/N$ 
(with ${\mathbf m}_i=(\cos\theta_i,\sin\theta_i)$),
an analytical BG canonical solution of this system predicts
a second-order phase transition from a low-energy ferromagnetic phase 
with magnetization $M\equiv|{\mathbf M}|\neq 0$, to a high-energy one with 
the spins homogeneously oriented on the unit circle and $M = 0$. 
The critical point is at energy density $u_c=0.75$,
and a caloric curve can be exactly obtained \cite{dauxois_01}.
On the other hand, it is quite simple to integrate numerically the 
equations of motion and to study the dynamical behavior of the system
by means of molecular dynamics simulations for a relatively large
number of spins $N$. 
When this is done for energy densities 
(slightly) below the critical point, with 
some nonzero-measure class of out-of-equilibrium initial conditions 
\cite{latora_02}, one finds 
that the system is dynamically stuck in quasistationary 
nonequilibrium states {\it whose duration diverges with} $N$.
Consistently with standard equilibrium statistical mechanics, 
a `dynamical temperature' $T$ can be defined as 
\begin{equation}
T(t)\equiv 2 K(t)/N,
\label{T}
\end{equation}
where $K$ is the kinetic energy and $t$ is time.
In the present approach, 
we use the qualification dynamical in the sense that this definition purely 
descends from dynamics and not from a thermal contact with a thermostat.
For system (\ref{H}), during the QSS, the temperature $T_{QSS}$ evolves into 
a first plateau {\it below} the BG equilibrium temperature $T_{BG}$, 
before relaxing to a second plateau that coincides with $T_{BG}$.
The difference between $T_{QSS}$ and $T_{BG}$ is maximal
for energy density around $u=0.69$.
$T_{QSS}$ depends on the size $N$
of the system, but tends to a well defined value $T_{\infty}\simeq0.38$ as 
$N\to\infty$.
Notice that other long-range 
interacting models display QSS with temperature {\it above} 
$T_{BG}$ \cite{nobre_01}.
Moreover, aging \cite{montemurro_01} and non-Gaussian one-body marginal 
velocity distributions characterize the QSS.
As strong numerical evidences indicate \cite{latora_02,nobre_01}, 
in the thermodynamic limit $N\to\infty$, the QSS lasts
forever. If we are attempting a thermodynamical description of
such a system, it is fundamental to discuss the zeroth principle under
these conditions.
Furthermore, we would like to know what would be the 
response of a thermometer testing the quasistationary temperature.

\begin{figure}
\begin{center}
\includegraphics[width=7cm,angle=0]{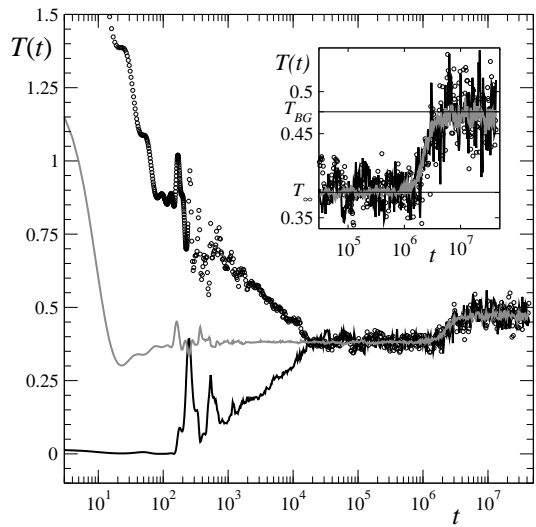}
\end{center}
\caption{\small
Temperature evolution of an isolated  
$N$-rotor system (Eq. (\ref{H})) in grey line,
and {\it cold} ({\it hot}) $M$-rotor subsystem
in black line (circles).
Inset: magnification of the crossover between $T_{QSS}$ and $T_{BG}$.
}
\label{fig1}
\end{figure}

In order to provide an answer to these questions
we numerically integrate the Hamilton equations using the $4$th order
symplectic Neri-Yoshida integrator \cite{yoshida} 
with energy conservation $\Delta E /E \simeq 10^{-4}$, 
under different setups.
In our first simulation we examine 
precisely the textbook construction for
deriving the canonical ensemble as a subsystem of the 
microcanonical ensemble.
First, we consider the isolated Hamiltonian system (\ref{H}),
composed by $N$ spins.
Our out-of-equilibrium initial conditions consist in setting 
every angle parallel, e.g., $\theta_i=0$, $\forall i$, and the angular momenta 
distributed inside an interval
around the origin with a fixed separation $a$ between them,  
but each with a random shift that could be up to $a$. 
We renormalize the distribution in order to have total energy 
density $u=0.69$ and zero total angular momentum.  
These initial conditions are very similar to the ones normally used in the literature, 
the so-called `\emph{waterbag}' initial conditions, but the present setup yields a lower initial 
temperature, just above $T_{\infty}$, which results in both smaller 
fluctuations and a longer duration of the anomalous plateau.  
Then, we take into account two distinct subsystems of the isolated system, 
each of them consisting of $M$ spins, with $M<<N$. 
The first (second) subsystem is composed by the most (least) energetic initial data, 
so that its initial temperature $T_M(0)=2K_M(0)/M$
is larger (smaller) than the one, $T_N(0)=2K_N(0)/N$, of the
isolated system as a whole.
Fig. \ref{fig1} shows the result of a {\it single} typical 
simulation, with $N=10^4$ and $M=5\times10^2$.
In both cases we see that $T_M$ relaxes after some time to $T_N$, 
{\it while the isolated system is still in the QSS}. 
This situation lasts until all systems undergo 
together the expected relaxation to $T_{BG}$, due to finite-size effects. 
It is important to stress that the subsystem starting with the higher temperature,
in its relaxation to the temperature of the isolated system,
crosses $T_{BG}$ with no signs of attempts to relax to $T_{BG}$ itself.
This result, for the QSS, precisely complies with the zeroth principle of equilibrium 
thermodynamics. Indeed, two systems are in thermal (meta)equilibrium
with a third system, and in thermal (meta)equilibrium with each other.
Also, this verification clearly opens the possibility for a generalized canonical treatment of the QSS,
as all subsystems with $M<<N$ share the same dynamical temperature after
an initial transient. 
Such a detailed study is presently under course \cite{baldovin_01}.

Our second simulation addresses the question of the measurability of the QSS 
dynamical temperature.
We do this by considering the system (\ref{H}) as a {\it thermostat} 
and by constructing a {\it thermometer} very different from the system 
(\ref{H}), in the sense that it complies with the usual BG laws. 
The difference lies, more precisely, in the fact that we choose {\it short-range} 
interactions for the thermometer, both for its own dynamics 
and for its coupling with the thermostat.
The thermometer is then composed by $M$ classical spins whose Hamiltonian is
\begin{equation}
H_{thermometer} = \sum_{j=1} ^{M} \frac{p_{j}^2}{2} + 
\sum_{j=1} ^{M}\left[1-cos(\theta_{j}-\theta_{j+1})\right].
\label{HM}
\end{equation}

It has the same potential as the thermostat, but  
it only connects first neighbors. 
This short-range interaction results in a standard BG system 
as we will see below.
The thermometer is prepared as follows: before entering into contact with 
the thermostat, we set the thermometer coordinates to $\theta_j=0$,  $\forall j$, 
and $p_j$ taken from a Gaussian distribution whose standard deviation is 
chosen in order to start with a particular temperature of our choice. 
We then let the system evolve freely for enough time until 
complete BG equilibrium is achieved with a Maxwellian one-body marginal 
velocity distribution. 
The $2M$ resulting coordinates are then used as initial conditions for the contact
with the thermostat. In this way we are confident that we are starting with a 
thermometer in a usual BG equilibrium.
On the other side, the thermostat is prepared in the `water bag' 
initial conditions previously described
and we let evolve the two systems separately, until any quick transient states 
have disappeared. 
At a convenient time $t_{contact}$ they are ``connected'' through an interaction
term
\begin{equation}
H_{int} = c \sum_{j=1} ^{M}
\left[1-cos(\theta_{j}-\theta_{\xi(j)})\right],
\label{Hint}
\end{equation}
where $\xi(i)$ is a random natural number between $1$ and $N$ (fixed once for ever) that describes 
the connection between thermometer and thermostat spins. 
We include also a coupling constant $c$ to regulate the intensity of the 
interaction term
(the coupling constant between spins of the same system 
is equal to unity).
If we call $H_{thermostat}$ the Hamiltonian (\ref{H}),  the total system after $t_{contact}$ is 
then described by the Hamiltonian
\begin{equation}
H=H_{thermostat}+H_{thermometer}+H_{int}.
\label{Htot}
\end{equation}

\begin{figure}
\begin{center}
\includegraphics[width=7cm,angle=0]{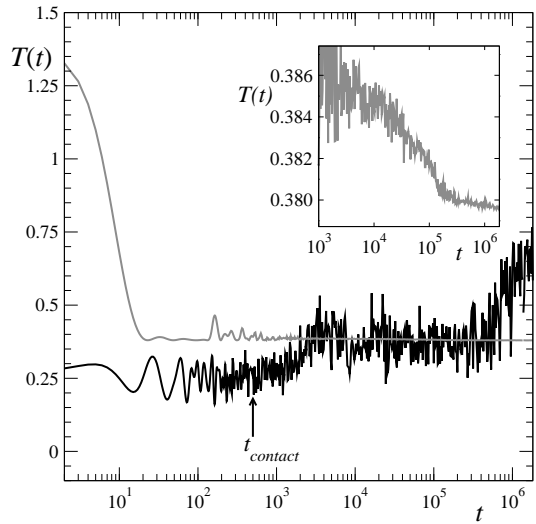}
\end{center}
\caption{\small
Temperature evolution of an $N$-rotor thermostat  
(Eq. (\ref{H})) in grey line,
and of an $M$-rotor thermometer (\mbox{Eq. (\ref{HM})}) in black line.
After $t_{contact}$ the systems interact through $H_{int}$.
Inset: magnification of the thermostat temperature minimum 
(see text for details).
}
\label{fig2}
\end{figure}

Results of a {\it single} typical simulation with $N=10^5$, 
$M=50$ and $c=5\times10^{-2}$ are shown in Fig. \ref{fig2}.
It is important to appropriately choose the range of the numerical value 
of the coupling constant $c$. Indeed, we want on one hand to establish a significative coupling between 
the systems, but on the other hand to produce a not too large perturbation of the thermostat. We expect this care to become less and less restrictive as we numerically approach the theoretical limit $(N,M,N/M)\to(\infty,\infty,\infty)$. 
We see that the thermometer temperature $T_M=2K_{thermometer}/M$, 
chosen in order to start below the thermostat temperature, 
stays few time steps in its initial equilibrium state and afterwards starts to 
grow rapidly to reach the thermostat temperature $T_N=2K_{thermostat}/N$, 
and {\it relaxation occurs completely within the QSS},
for $\Delta t\approx10^5$ time steps 
(fluctuations are of course present because of finite-size effects). 
Differently with the previous case, the thermometer eventually 
begins to relax to its equilibrium temperature, {\it before} the thermostat 
starts its final thermalization. 
As explained below, we consider this as one more finite-size effect.
Note in the inset of Fig. \ref{fig2} that the time at which the 
thermometer leaves the thermostat temperature approximately coincides 
with its minimum, that is known to be present just after 
the thermostat finally relaxes to $T_{BG}$ \cite{latora_02}.

When the thermometer is prepared in order to have, before contact, a temperature which is 
higher than that of the thermostat, our results show no clear 
signs of thermalization. 
$T_M$ increases steadily  until it attains its definite 
equilibrium. This behavior will hopefully disappear when computers will allow simulations with even larger systems. 
It is also interesting to notice that, even preparing the thermometer 
with a temperature below $T_\infty=0.38$, thermalization occurs only 
for $N$ and $N/M$ sufficiently large. 
For example, simulations with $N=5\times10^5$ and $M=5\times10^2$ 
do not show any thermalization.

Taking all these facts in consideration, namely, thermometer temperature 
reaching BG equilibrium before the thermostat, no relaxation 
for $T_{M}(t_{contact})>T_{N}(t_{contact})$, no relaxation for too small 
$N$ and $N/M$, and also taking into account the fact that the system is aging, one may suspect what follows.  
The model (\ref{H}) behaves like having 
some internal mechanism that, after a certain
amount of time, for finite $N$, 
pulls the system out of the QSS to a BG equilibrium. 
This mechanism works like an internal clock that regulates this 
thermalization time, and may function like a potential well whose 
deepness decreases with time. 
A system with large enough fluctuations, compared to the well depth, 
will never be confined by the potential well. 
A system with small enough fluctuations is constrained to the well, 
but just for a limited amount of time, until the well becomes shallow and 
its depth becomes comparable to the fluctuations. 
In our case, the effect of the well would be to restrict the system 
to visit just a part of the phase space, 
while not being in any well during a long time would be associated to an 
homogeneous visit, the system thus becoming ergodic and relaxing 
to the expected BG temperature (see \cite{baldovin_02} for a low-dimensional analogy). 
Consequently, within this picture, fluctuations would greatly influence 
the system permanence in the QSS. 
Note that this scenario is also consistent with the 
relaxation of subsystems of an isolated system as observed in Fig. \ref{fig1}.

In order to state our conclusions, let us recall that in the last years
considerable interest has been raised in the study of long-range
interacting systems \cite{dauxois_02}. 
Significant progress has been made in the description and comprehension
of out-of-equilibrium QSS that by no means accommodate within the BG 
scenario.
These QSS display an anomalous temperature
plateau \cite{latora_02} and  various other anomalies that 
are consistent with the nonextensive statistical mechanics \cite{tsallis_02} 
picture. 
Among them, vanishing Lyapunov exponents \cite{anteneodo_01} 
(see \cite{baldovin_03} for illustrative connections between nonextensive
statistical mechanics and vanishing Lyapunov exponents),
L\'evy walks and anomalous diffusion \cite{latora_01},  
aging involving $q$-exponential (i.e., asymptotically power-law) 
time correlation functions \cite{montemurro_01}, $q$-exponential one-body marginal 
velocity distributions \cite{latora_02}, and 
$q$-exponential relaxation to the BG equilibrium \cite{tsallis_01}.
Summarizing, if one wants to design a generalized
statistical mechanical approach for the description of these QSS,
a fundamental step concerns the validity of the zeroth
principle.
Our present results exhibit 
that this basic law of thermodynamics also applies to the QSS, and has therefore a domain of  
validity which is wider that the one normally associated with it within BG statistical mechanics. 
Furthermore, we have provided evidence that the QSS dynamical temperature
is actually detectable by a normal short-range thermometer.
We believe that the present findings establish fundamental grounds for forthcoming research in
the area.

\section*{Acknowledgments}
We thank C. Anteneodo for
useful remarks and discussions.
We have benefitted from partial support by CNPq, CAPES, PRONEX and 
FAPERJ (Brazilian agencies).

\vspace{2cm}
\vspace{2cm}

\end{multicols}
\end{document}